\documentstyle[12pt,epsf]{article}
\evensidemargin\oddsidemargin
\newdimen\unit
\def\point#1 #2 #3{\vbox to0pt{\kern-#2\unit
  \hbox{\kern#1\unit$#3$}\vss}
 \nointerlineskip}
\def\btorho{\bar B^0\to\rho^+ l^- \bar\nu_l}
\def\btopi{\bar B^0\to\pi^+ l^- \bar\nu_l}
\def\btokstargamma{\bar B\to K^* \gamma}
\def\vub{|V_{ub}|}
\def\qsqmax{q^2_{\rm max}}
\def\w{\omega}
\def\azero{A_{0,3}(0)}
\def\gev{\,{\rm Ge\kern-0.1em V}}
\def\mev{\,{\rm Me\kern-0.1em V}}
\makeatletter
\def\big#1{{\hbox{$\left#1\vbox to1.012\ht\strutbox{}\right.\n@space$}}}
\def\Big#1{{\hbox{$\left#1\vbox to1.369\ht\strutbox{}\right.\n@space$}}}
\def\bigg#1{{\hbox{$\left#1\vbox to1.726\ht\strutbox{}\right.\n@space$}}}
\def\Bigg#1{{\hbox{$\left#1\vbox
to2.083\ht\strutbox{}\right.\n@space$}}}
\long\def\@makecaption#1#2{
 \vskip 15pt 
 \small
 \setbox\@tempboxa\hbox{{\bf #1} \ #2}
 \ifdim \wd\@tempboxa >\hsize \unhbox\@tempboxa\par \else \hbox
to\hsize{\hfil\box\@tempboxa\hfil} 
 \fi}
\def\er#1#2{\relax\ifmmode{}^{+#1}_{-#2}\else$^{+#1}_{-#2}$\fi}
\def~{\ifmmode\phantom{0}\else\penalty10000\ \fi}
\makeatother
\def\tstrut{\vrule height2.5ex depth0pt width0pt} 

\begin{document}
\def\Atwovalue{0.28\er96\er45}
\begin{titlepage}
\begin{flushright}
Granada Preprint UG--DFM--3/96\\
Southampton Preprint SHEP 96--01\\
hep-ph/9602201
\end{flushright}

\bigskip

\begin{center}
{\Huge First Lattice Study of the Form Factors $A_0$ and $A_3$ in the 
Decay $\bar B^0\to\rho^+ l^- \bar\nu_l$.}\\[5em]
{\large\it UKQCD Collaboration}\\[1.5ex]
{\bf J M Flynn}\\
Department of Physics, University of Southampton, Southampton SO17
1BJ, UK\\[0.5ex]
{\bf J Nieves}\\
Departamento de Fisica Moderna, Avenida Fuentenueva, 18071 Granada,
Spain
\end{center}

\bigskip

\begin{abstract}
We report on a lattice calculation of the form factors $A_0$ and $A_3$
for the pseudoscalar to vector meson semileptonic decay $\btorho$. We
find that resonant (or pole-type) contributions alone are unable to
describe these two form factors simultaneously. For the quantity
$A_0(q^2{=}0)$, which is important phenomenologically for the
determination of $\vub$, we extract a range of values, $A_0(q^2{=}0) =
(0.16$--$0.35)\er96$, where the range is due to systematic uncertainty
and the quoted error is statistical. We have also determined
$A_2(q^2{=}0) = \Atwovalue$.
\end{abstract}

\end{titlepage}

\section*{Introduction}

The semileptonic decay $\btorho$ is determined by the matrix element
of the $V{-}A$ weak current between a $\bar B$ meson and a $\rho$
meson. The matrix element is,
\begin{equation}
\langle \rho(k,\eta)| \bar u \gamma_\mu(1-\gamma_5)b| \bar B(p) \rangle
  = \eta^{*\beta} T_{\mu \beta}, \label{eq:btorho1}
\end{equation}
with form factor decomposition,
\begin{eqnarray}
T_{\mu \beta} &=& \frac{2V(q^2)}{m_B+m_\rho} 
\epsilon_{\mu \gamma \delta \beta}p^{\gamma}k^{\delta} 
- i(m_B+m_\rho)A_1(q^2)g_{\mu \beta} \nonumber \\
& & \mbox{} + i \frac{A_2(q^2)}{m_B+m_\rho} (p+k)_{\mu}q_{\beta} - 
i \frac{A(q^2)}{q^2}2m_\rho q_{\mu}(p+k)_{\beta},  \label{eq:btorho2}  
\end{eqnarray}
where $q = p - k$ is the four-momentum transfer and $\eta$ is the
$\rho$ polarisation vector. The form factor $A$ can be written as
\begin{equation}
A(q^2) = A_0(q^2)-A_3(q^2),
\end{equation}
where,
\begin{equation}\label{eq:a3def}
A_3(q^2) = \frac{m_B+m_\rho}{2m_\rho}A_1(q^2)- 
\frac{m_B-m_\rho}{2m_\rho}A_2(q^2), 
\end{equation}
with $A_0(0)=A_3(0)$. In the limit of zero lepton masses, the term
proportional to $A$ in equation~(\ref{eq:btorho2}) does not contribute
to the total amplitude and hence to the decay rates. Pole dominance
models suggest that $V$, $A_i$ for $i=1,2,3$ and $A_0$ correspond to
$1^-$, $1^+$ and $0^-$ exchanges respectively in the
$t$-channel~\cite{bsw}.

Neglecting corrections suppressed by inverse powers of the heavy quark
mass $M$, the following relations hold when $q^2$ is close to the
maximum recoil value $\qsqmax = (M - m_\rho)^2$~\cite{isgurwise:hqet}
\begin{equation}\label{eq:hqs-scaling-a0a3}
A_0\Theta /M^{1/2} = {\rm const}, \qquad
A_3\Theta /M^{3/2} = {\rm const},
\end{equation}
where $\Theta$ arises from perturbative corrections and is chosen to
be $1$ at the $B$ mass. In leading order~\cite{neubert:physrep},
\begin{equation}\label{eq:theta}
\Theta = \Theta(M,m_B) = \left( \frac{\alpha_s(M)}{\alpha_s(m_B)} 
                         \right)^{\frac{2}{\beta_0}}.
\end{equation}
In the calculations reported below, we will use $\beta_0 = 11$ in the
quenched approximation and $\Lambda_{\rm QCD} = 200\mev$.

The differential decay rate $d\Gamma(\btorho)/dq^2$ at $q^2=0$ is
given by\footnote{The decay rate expression in
equation~(\protect\ref{eq:diffdecayrate}) is naturally expressed in
terms of $\big|A_3(0)\big|$, but since our lattice data for $A_0$ are
of much higher quality, we prefer to use the relation $A_0(0) = A_3(0)$
and express the result in terms of $\big|A_0(0)\big|$},
\begin{equation}\label{eq:diffdecayrate}
{d\Gamma(\btorho)\over dq^2}\Big|_{q^2=0} = 
 {G_F^2 \vub^2 \over 192 \pi^3 m_B^3}\, (m_B^2 - m_\rho^2)^3 
 \big| A_0(0) \big|^2,
\end{equation}
and is determined by the Cabibbo--Kobayashi--Maskawa (CKM) mixing
matrix element $\vub$ and the form factor $A_0(q^2{=}0)$, which
accounts for the nonperturbative QCD contributions.  A lattice
measurement of $A_0(0)$ can therefore be combined with experimental
measurements of the differential decay rate to extract $\vub$.

It has previously been suggested~\cite{odonnell,santorelli} that
hadronic uncertainties can be reduced by combining a measurement of
the differential rate with the recent
measurement~\cite{cleo:btokstargamma} of the rare radiative decay
$\btokstargamma$ to determine $\vub$, according to the relation:
\begin{equation}\label{eq:Roverdiffrate}
{R(\btokstargamma) \over d\Gamma(\btorho)/dq^2|_{q^2=0}} =
  {192\pi^3\over G_F^2} {1\over\vub^2}
  {(m_B^2-m_{K^*}^2)^3\over(m_B^2-m_\rho^2)^3}
  {1\over m_b^3}
  \left|{2 T_2^{B\to K*}(0)\over A_0^{B\to\rho}(0)}\right|^2.
\end{equation}
Here, the hadronisation ratio $R(\btokstargamma) =
\Gamma(\btokstargamma)/\Gamma(b\to s\gamma)$ depends only on the form
factors $T_1(0) = iT_2(0)$ for the exclusive radiative decay, and is
given up to ${\cal O}(1/m_b^2)$ and perturbative QCD corrections
by,\footnote{The theoretical prediction for this ratio may be subject
to long-distance effects~\cite{LD:atwoodetal,longdist}}
\begin{equation}
R(\btokstargamma) = 4 \left(m_B\over m_b\right)^3
              \left(1 - {m_{K^*}^2\over m_B^2}\right)^3
              \big| T_2(0) \big|^2.
\end{equation}
Using equation~(\ref{eq:Roverdiffrate}) to determine $\vub$ is
advantageous if the ratio $2T_2^{B\to K^*}(0)/A_0^{B\to\rho}(0)$ is
known in advance.
There are some predictions that this ratio is close to
$1$~\cite{odonnell,sumrules:cds}. Lattice studies can help here by
allowing a model-independent calculation.

QCD sum rule analyses~\cite{sumrules:cds}
--\nocite{sumrules:abs,sumrules:ball}
\nocite{sumrules:bbd,sumrules:bkr,sumrules:cs} \cite{sumrules:cdp},
quark models~\cite{stech,cybernetics} and a combined theoretical and
phenomenological analysis~\cite{alop} give the following picture for
the form factors for $\bar B \to \rho$ transitions: $V$ and $A_0$ have
significant $q^2$ dependence, while $A_1$ is rather flat in $q^2$ and
$A_2$ shows intermediate behaviour.

In this note, we present the first lattice study of the form factors
$A_0(q^2)$ and $A_3(q^2)$, and give a range of values for $A_0(q^2=0)$
which, given experimental measurements of
$d\Gamma(\btorho)/dq^2|_{q^2=0}$, may be used to determine $\vub$,
according to equation~(\ref{eq:diffdecayrate})\footnote{The CLEO
collaboration have measured exclusive semileptonic $B\to\rho$ and
$B\to\pi$ decays, and have preliminary results for $q^2$
distributions~\protect\cite{cleo:exclusive}.}. We first study
$A_0(q^2)$ and then argue that a combined fit to $A_0$ and $A_3$ can
give more accurate results for $A_0(0)$ as lattice data improve. We
also report on a value for $A_2(q^2{=}0)$. Other lattice calculations
of $A_2$ for $\btorho$ can be found in
references~\cite{ape:btopi,elc:btopi,wup:slff}.

\section*{Lattice Details}

The results described below come from $60$ $SU(3)$ gauge
configurations generated by the UKQCD collaboration on a $24^3 \times
48$ lattice at $\beta = 6.2$ in the quenched approximation. The ${\cal
O}(a)$ improved Sheikholeslami--Wohlert (SW)~\cite{sw-action} action
was used for fermions, with ``rotated'' fermion fields appearing in
all operators used for correlation function
calculations~\cite{heatlie:clover-action}. 

Three-point correlators of the operator in equation~(\ref{eq:btorho1})
with a heavy pseudoscalar meson (the ``$\bar B$'') and a light vector
meson were calculated, for various combinations of heavy ($\bf p$) and
light ($\bf k$) meson three momenta. We refer to each combination with
the notation $|{\bf p}|\to |{\bf k}|$ where the momenta are in lattice
units of $\pi/12a$ (a subscript $\perp$ on $\bf k$ indicates that $\bf
p$ and $\bf k$ are perpendicular). Here, the inverse lattice spacing
determined from the $\rho$ mass is $a^{-1} =
2.7(1)\gev$~\protect\cite{ukqcd:strange-prd}. We use this value for
$a^{-1}$ since we are dealing with decays to a $\rho$ meson, but it
should be remembered that different physical observables lead to
different values for $a^{-1}$ (UKQCD determinations of $a^{-1}$ are
summarised in~\cite{ukqcd:static}). This uncertainty in $a^{-1}$
should have little effect on the dimensionless form factors evaluated
in this paper.

For relating lattice results to the continuum, we have used the
perturbative value for the renormalisation constant of the axial
current operator~\cite{zva}:
\begin{equation}
Z_A =  0.97.
\end{equation}

Uncorrelated fits were used for extrapolations in the heavy quark
mass, although the extraction of the form factors from calculated
three-point correlation functions used correlated
fits~\cite{ukqcd:dtok}. Statistical errors are 68\%
confidence limits obtained from $250$ bootstrap samples: unless
otherwise specified, errors quoted below are statistical only.  More
details of the lattice calculations can be found in
references~\cite{ukqcd:hlff} and~\cite{ukqcd:btorho}.

The SW improved action reduces the leading discretisation errors from
${\rm O}(a)$ in the Wilson fermion action to ${\rm O}(\alpha_s
a)$. For quark masses $m$ around that of the charm quark, $\alpha_s m
a$ can be of order 10\%. For this first study we have not tried to
disentangle these $m$-dependent discretisation effects from physical
mass dependence. We leave them as a component of the systematic error.

\section*{Computing $A_0(0)$}

We have extracted values for $A_0(q^2)$ for five momentum channels (we
cannot extract $A_0$ for the zero recoil channel, which has $q^2 =
\qsqmax$), and for four heavy quark masses around the charm mass. We
choose channels where the value of $\w$ defined by,
\begin{equation}
\w = v\cdot v' = {M^2 + m^2 - q^2 \over 2 M m},
\end{equation}
is nearly constant as the heavy quark mass varies, where $M$, $v$ and
$m$, $v'$ are the masses and four-velocities of the pseudoscalar and
vector meson respectively. For these channels, we have extrapolated
linearly and quadratically in $1/M$ to the $B$ scale at constant $\w$,
by appealing to heavy quark symmetry. This gives five values for the
desired heavy-to-light form factor $A_0(q^2)$ with $q^2$ values near
$\qsqmax$: they are shown in table~\ref{tab:a0a3qsq} (the table also
gives values for $A_3(q^2)$ for later use).
\begin{table}
\hbox to\hsize{\def\arraystretch{1.2}\hss\begin{tabular}{lcccc}
\hline
\tstrut
channel & $q^2/\!\gev^2$ & $A_0(q^2)$ & $A_3(q^2)$ &
  $A_3^{\rm calc}(q^2)$\\[0.5ex]
\hline
\tstrut
$1\to{\sqrt2}_\perp$
        &  14.4\er33 &  0.98\er{12}{13}  & $-0.51\er{45}{61}$
& $-0.50\er{54}{72}$ \\
$0\to\sqrt2$
        &  15.3\er33 &  0.94\er{8~}{8~}  & $-0.52\er{23}{31}$
& $-0.64\er{31}{42}$ \\
$1\to1_\perp$
        &  16.7\er22 &  1.26\er{17}{12}  & $-0.34\er{59}{65}$
& $-0.25\er{67}{68}$ \\
$0\to1$ &  17.5\er22 &  1.38\er{11}{10}  & $-0.60\er{49}{44}$
& $-0.64\er{59}{53}$ \\
$1\to0$ &  19.71\er11 &  2.24\er{55}{61}  & --- & --- \\
\hline
\end{tabular}\hss}
\caption[]{Values of the form factors $A_0(q^2)$ and $A_3(q^2)$ for
$\btorho$.  Errors are statistical only. The final column contains
$A_3$ values obtained from previous results for $A_1$ and
$A_2$~\protect\cite{ukqcd:btorho}.}
\label{tab:a0a3qsq}
\end{table}

These values can then be fitted to various $q^2$ dependences for $A_0$
and an extrapolation made to determine $A_0(0)$. We will refer to this
as the ``constant $\w$'' method below. As a check, we also extrapolate
to $q^2=0$ for each of our four heavy quark masses around the charm
mass, and then extrapolate $A_0(0)$ to the $B$ scale: this will be
referred to as the ``$q^2=0$ extrapolation'' method. Our methods are
described in more detail in references~\cite{ukqcd:hlff}
and~\cite{ukqcd:btorho}.

\begin{figure}
\hbox to\hsize{\epsfxsize=0.48\hsize
\epsffile[30 43 513 507]{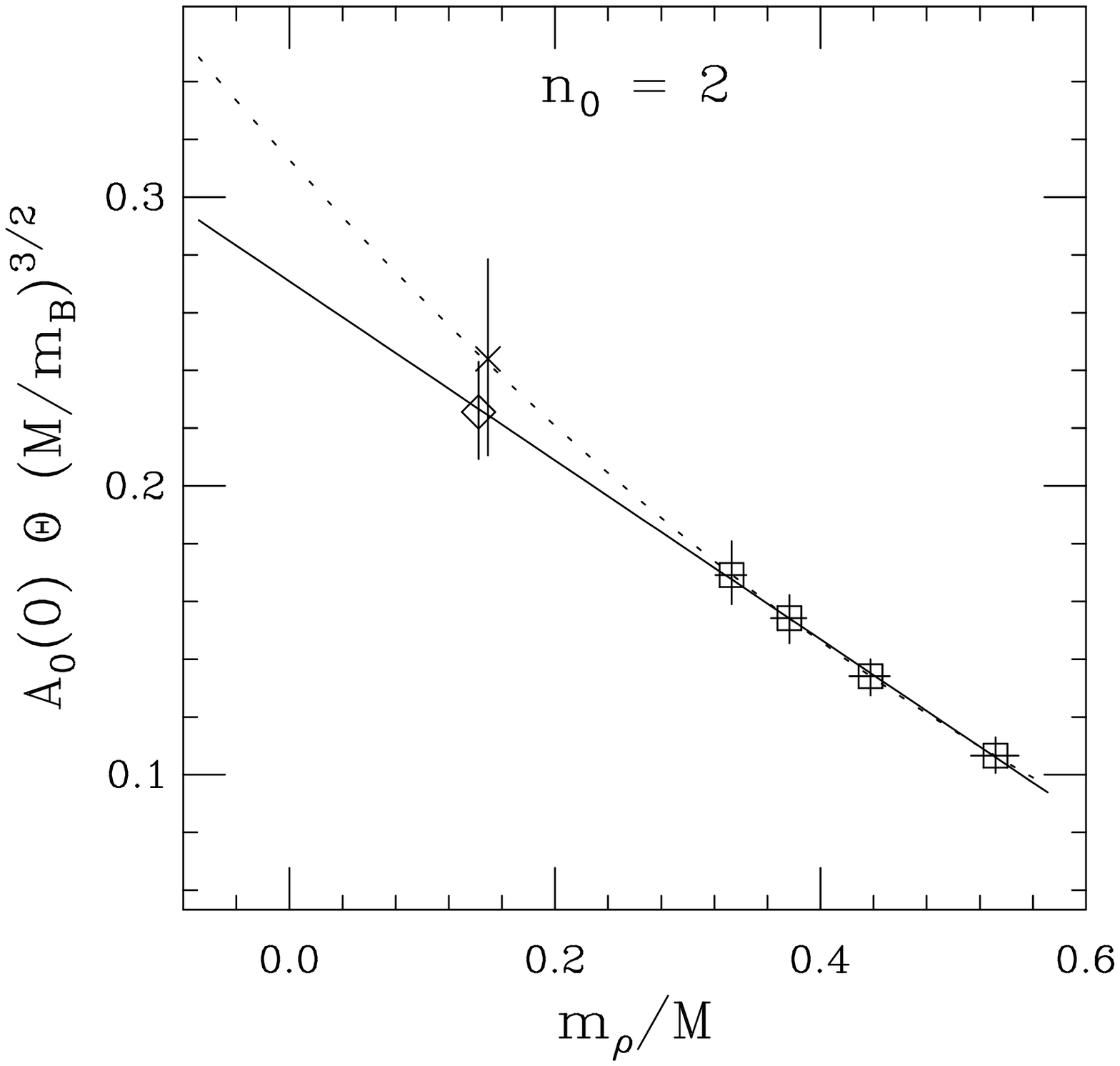}\hfill
\epsfxsize=0.48\hsize
\epsffile[30 43 513 507]{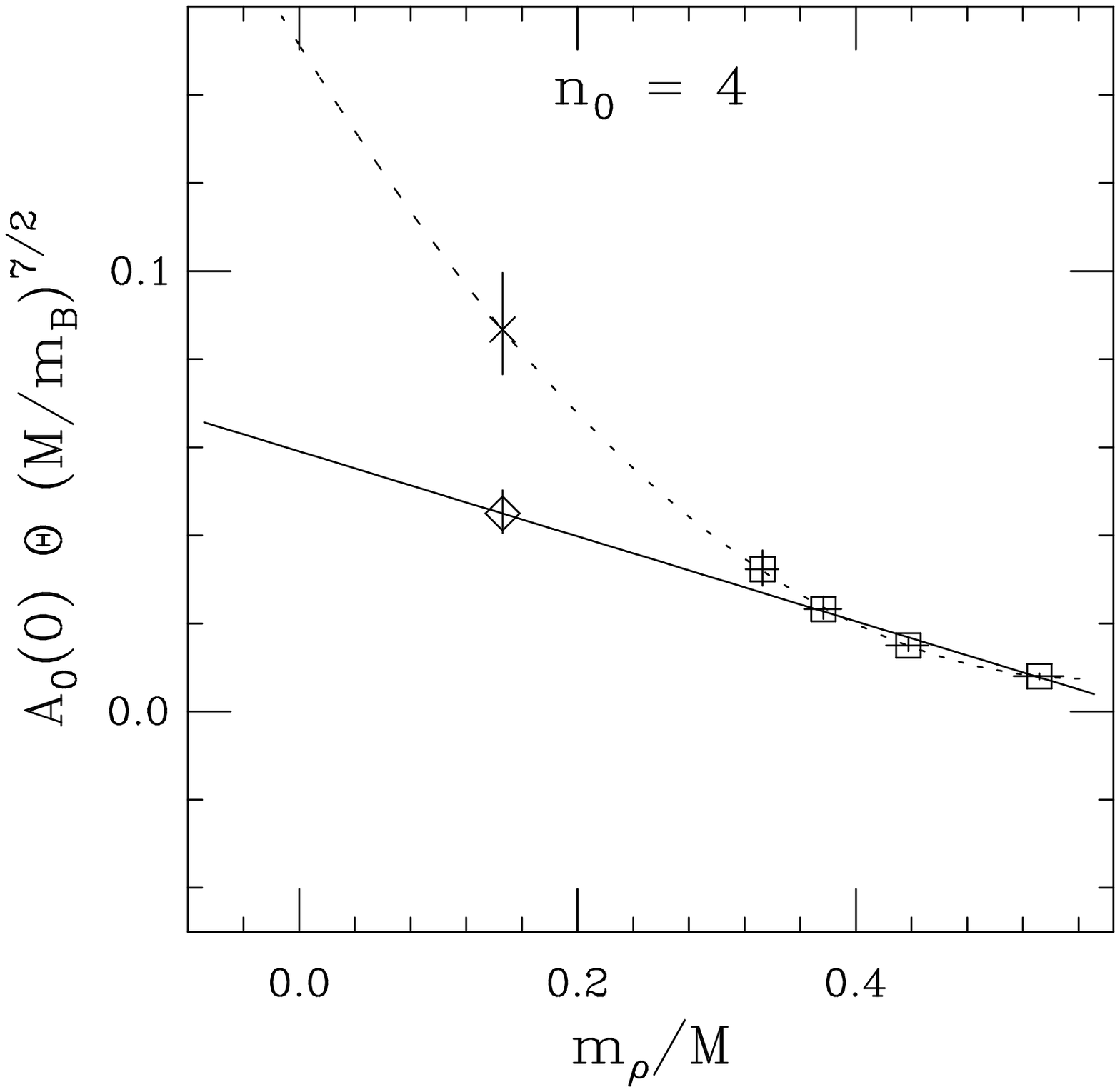}}
\caption[]{Comparison of linear and quadratic in $1/M$ extrapolations
of $A_0(0)$ (``$q^2=0$ extrapolation'' method). $A_0(q^2)$ has been
fitted at each heavy mass value. The values of $A_0(0)$ from the
individual fits (squares) are then extrapolated linearly (solid curve)
or quadratically (dashed curve) in $1/M$ to the $B$ scale (diamond or
cross). In the left hand plot, the extrapolated points have been
displaced slightly for clarity. On the left a dipole form [$n_0=2$ in
equation~(\protect\ref{eq:a0poleforms})] has been used at each heavy
mass, and on the right a quadrupole ($n_0=4$) form.}
\label{fig:a0-qsqzero-disagree}
\end{figure}

We have fitted $A_0(q^2)$ to $q^2$ dependences of the type,
\begin{equation}\label{eq:a0poleforms}
A_0(q^2) = {A_0(0)\over (1 - q^2/m_0^2)^{n_0}},
\end{equation}
for a range of ``polar powers'' $n_0 = 1,\ldots,5$. The higher powers,
while not physically motivated, allow us to explore different $q^2$
dependences away from the pole region. The $q^2$ dependence of $A_0$
in equation~(\ref{eq:a0poleforms}) is consistent with the heavy quark
symmetry requirement of equation~(\ref{eq:hqs-scaling-a0a3}), which
applies for $q^2$ close to $\qsqmax$, provided that,
\begin{equation}
A_0(0) \sim M^{-n_0 + 1/2}.
\end{equation}
This dependence on the heavy mass has been used when scaling $A_0(0)$
values from the charm mass to the bottom mass in the $q^2=0$
extrapolation method (see figure~\ref{fig:a0-qsqzero-disagree}).  All
extrapolations in the heavy pseudoscalar mass $M$ have been performed
to ${\cal O}(1/M^2)$. The results for both methods are given in
table~\ref{tab:a0only-qsqzero}.
\begin{table}
\hbox to\hsize{\def\arraystretch{1.2}\hss\begin{tabular}{c|c|ccc}
\hline
\tstrut
      & $q^2{=}0$ extrapolation
 & \multicolumn{3}{c}{constant $\w$} \\[-0.7ex]
$n_0$ & $A_0(0)$ & $A_0(0)$ & $m_0/\!\gev$ & $\chi^2/{\rm dof}$
 \\[0.5ex]
\hline
\tstrut
1 & 0.35\er56 & 0.34\er76 & 4.8\er21 & 0.32 \\
2 & 0.24\er33 & 0.23\er86 & 5.4\er43 & 0.39 \\
3 & 0.16\er22 & 0.19\er86 & 6.0\er54 & 0.42 \\
4 & 0.09\er11 & 0.17\er96 & 6.5\er65 & 0.44 \\
5 & 0.04\er11 & 0.16\er96 & 7.0\er75 & 0.45 \\
\hline
\end{tabular}\hss}
\caption[]{Values of the form factor $A_0(0)$ and the mass parameter
$m_0$ for $\btorho$ for different assumed $q^2$ dependences given in
equation~(\protect\ref{eq:a0poleforms}). Errors are statistical only.}
\label{tab:a0only-qsqzero}
\end{table}

\begin{figure}
\hbox to\hsize{\hfill\epsfxsize=0.5\hsize
\epsffile[52 35 513 507]{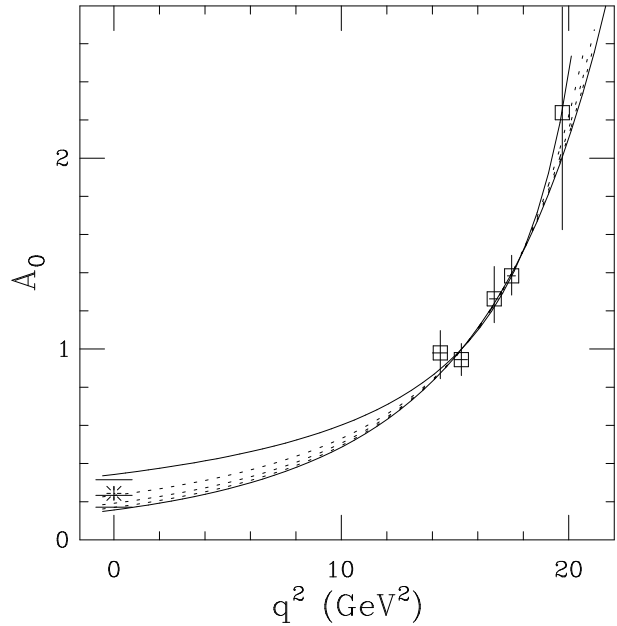}\hfill}
\caption[]{$A_0(q^2)$ for the decay $\btorho$ obtained by the constant
$\w$ method. Squares denote points extrapolated from measurements
around the charm scale in channels where $\w$ is constant (or nearly
constant). The curves shown are fits to
equation~(\ref{eq:a0poleforms}) with values $n_0 = 1,5$ (solid) and
$n_0 = 2,3,4$ (dotted) for the polar power. The three horizontal bars
at $q^2=0$ show the value and errors for $A_0(0)$ when $n_0 = 2$, and
the burst point shows the value obtained from the $q^2=0$
extrapolation method, also with $n_0 = 2$. For both methods, the value
of $A_0(0)$ decreases as $n_0$ increases.}
\label{fig:a0only-const-omega}
\end{figure}

Table~\ref{tab:a0only-qsqzero} shows that the $q^2=0$ extrapolation
method gives results compatible with the constant $\w$ method for
powers $1$, $2$ and $3$ in the fitted $q^2$ dependence, but that the
agreement gets worse for higher powers. This discrepancy is not
significant however, because the $1/M$ extrapolations are not under
control for higher polar powers. Indeed, for the lower powers the
linear and quadratic $1/M$ extrapolations agree, but for the higher
powers they do not, indicating that higher $1/M$ corrections must be
included. This point is illustrated in
figure~\ref{fig:a0-qsqzero-disagree}.  For the constant $\w$ method,
in contrast, the linear and quadratic $1/M$ extrapolations all agree
within errors for each momentum channel studied. Our results for fits
to $A_0(q^2)$ using the extrapolated points at the $B$ scale are shown
in figure~\ref{fig:a0only-const-omega}.

In summary, we find that there is some uncertainty in the
extrapolation to $q^2=0$ (in contrast to our finding for $A_1$ in our
earlier studies~\cite{ukqcd:btorho}). The uncertainty is smaller for
the constant $\w$ procedure, but nonetheless, a broad range of
$0.16$--$0.34$ has to be given for $A_0(0)$. Higher powers of $n_0$ do
not increase this range when $A_0(0)$ is extracted using the constant
$\w$ method. We discard the results with $n_0 \geq 4$ obtained from
the $q^2=0$ extrapolation method.

\section*{Constrained Fits to $A_0$ and $A_3$}

To gain more control over the extrapolations necessary to determine
$A_0(0)$, we combine them with a determination of $A_3(0)$ and apply
the constraint $A_0(0) = A_3(0)$. We have previously used this idea to
perform constrained fits to $f^+$ and $f^0$ in $\btopi$ as well as
$T_1$ and $T_2$ for $\btokstargamma$. To fit both $A_0$ and $A_3$ with
``resonant'' (pole or multipole) behaviour in $q^2$, whilst
maintaining the constraint $A_0(0) = A_3(0)$ and agreement with the
heavy quark symmetry requirements of
equation~(\ref{eq:hqs-scaling-a0a3}) at $\qsqmax$, would require a
model of the form,
\begin{eqnarray}
A_0(q^2) &=& {\azero\over (1-q^2/m_0^2)^{n_0}}
  \label{eq:a0poleform2},\\
A_3(q^2) &=& {\azero\over (1-q^2/m_3^2)^{n_0+1}}
  \label{eq:a3poleform}.
\end{eqnarray}
We find that such a model cannot describe our data.

We were able to extract $A_3(q^2)$ at the $B$ scale for four momentum
channels in our lattice simulation. The channel with the largest value
of $q^2$, $1\to0$ (the zero recoil channel, $0\to0$ cannot be
extracted directly for $A_3$), was too noisy to be extracted reliably:
as we shall see, better quality lattice data allowing its extraction
would be the easiest way to improve our results. Our results for
$A_3(q^2)$ after extrapolation to the $B$ scale in the constant $\w$
method are included in table~\ref{tab:a0a3qsq} (fourth column of
table).

\begin{figure}
\hbox to\hsize{\epsfxsize=0.48\hsize
\epsffile[30 43 513 507]{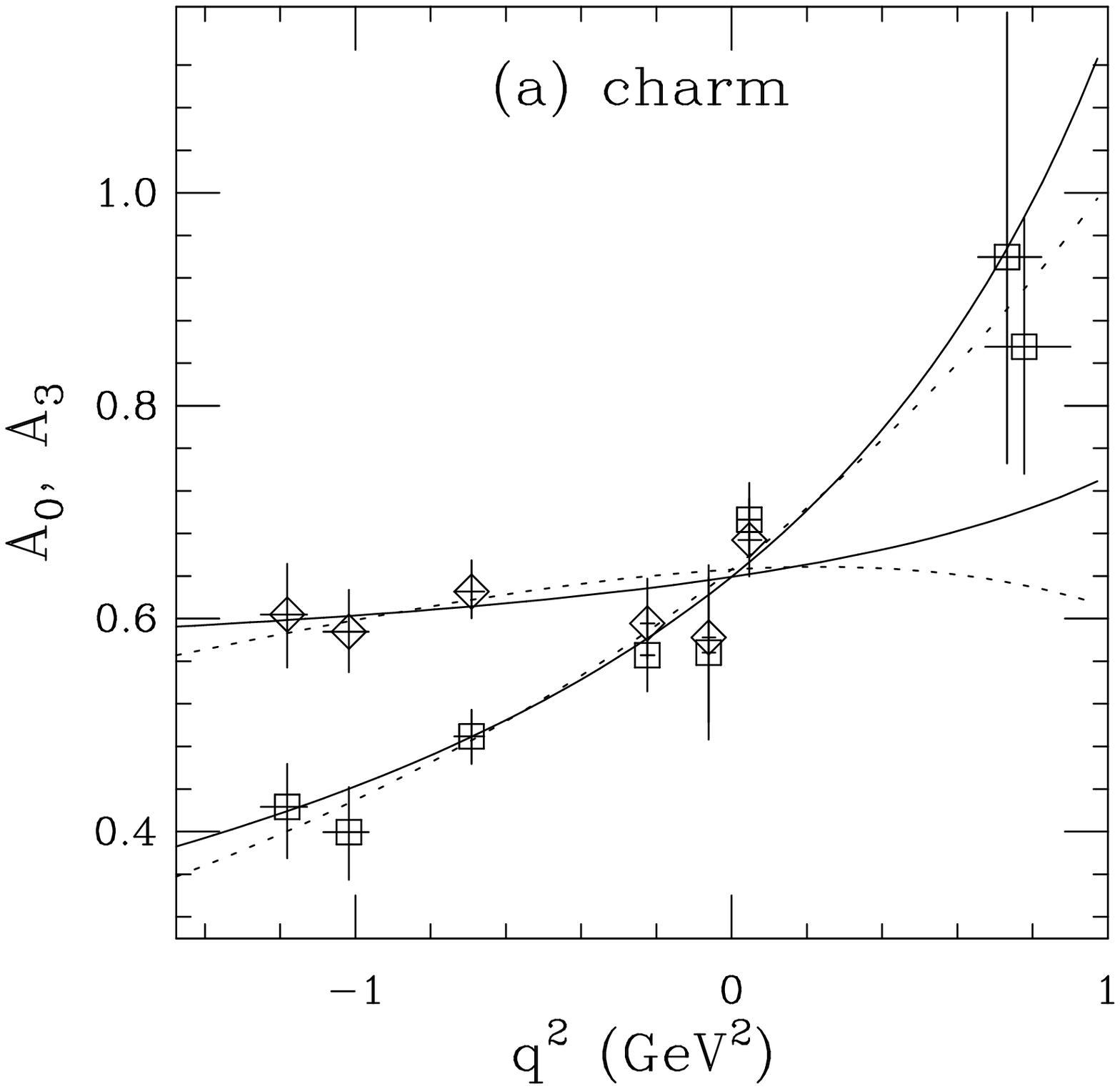}\hfill
\epsfxsize=0.48\hsize
\epsffile[30 43 513 507]{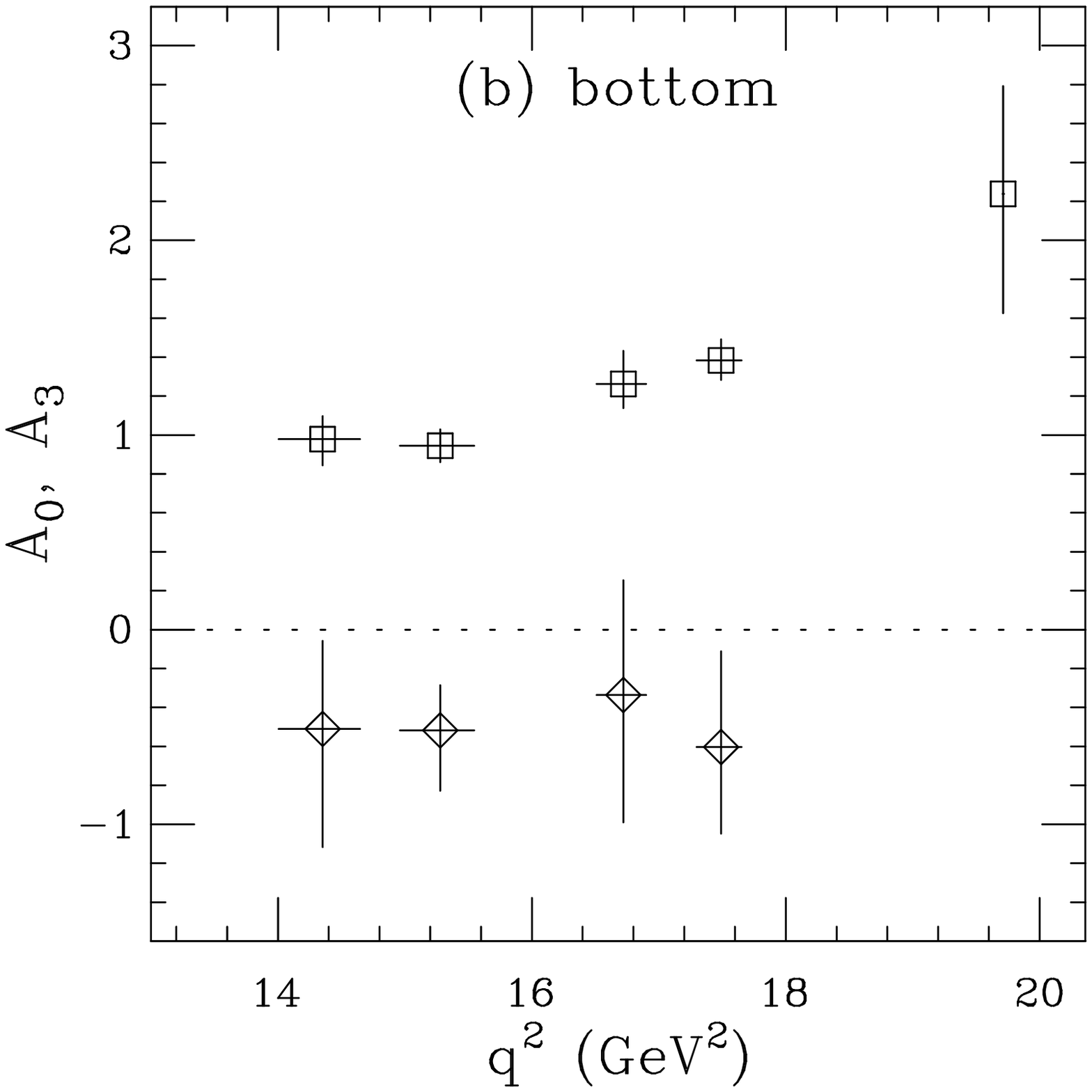}}
\caption[]{Comparison of $q^2$ dependences of $A_0$ and $A_3$ at (a)
the charm (left) and (b) the bottom (right) scales. The squares are
values for $A_0$ and the diamonds are values for $A_3$. The charm plot
in (a) uses our data with $\kappa=0.129$ for the heavy quark and with
the light quarks extrapolated to $\kappa_{\rm crit} =
0.14315(1)$~\protect\cite{ukqcd:strange-prd}. The curves on the left
hand plot are fits using the model of
equations~(\protect\ref{eq:a0model}) and~(\protect\ref{eq:a3model}),
with $n_0 = 1$ (solid) and $n_0 = 5$ (dotted), as explained in the
text.}
\label{fig:a0a3-c-and-b}
\end{figure}

We find that $A_3$ is positive for $q^2$ close to $\qsqmax$ in our
measured data, with heavy quark masses around the charm
scale. However, once extrapolated to the bottom mass, $A_3(q^2)$ is
{\em negative\/} for $q^2$ close to $\qsqmax$. $A_3$ is a linear
combination of $A_1$ and $A_2$, with positive and negative
coefficients respectively, where the $A_1$ contribution is most
important at the charm scale, but $A_1$ and $A_2$ contribute more
equally at the bottom scale. Moreover, heavy quark symmetry predicts
that $A_1$ decreases with increasing heavy quark mass, while $A_2$
increases. In figure~\ref{fig:a0a3-c-and-b} we show the $q^2$
dependence of $A_0$ and $A_3$ at the two scales (the meaning of the
curves in the charm-scale plot is explained later).

\begin{figure}
\hbox to\hsize{\hfill\epsfxsize=0.5\hsize
\epsffile[30 43 513 507]{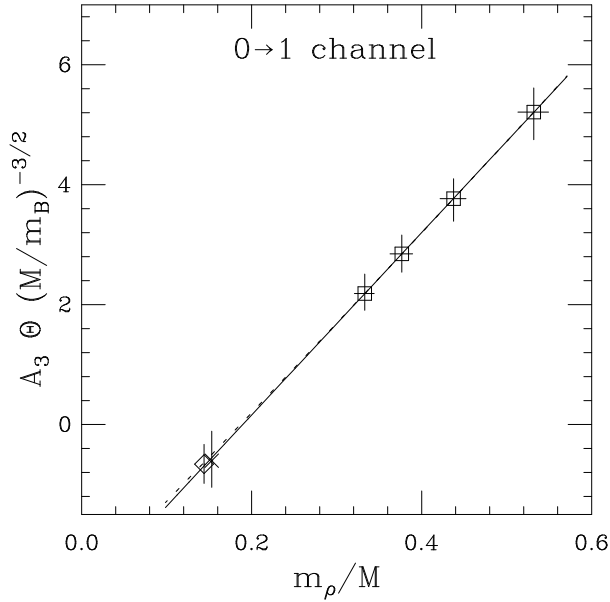}\hfill}
\caption[]{$1/M$ extrapolation for $A_3$ for the $0\to1$ channel, from
the charm scale to the bottom scale. The squares are measured points
for our four heavy mass values, which are extrapolated linearly (solid
curve) or quadratically (dashed curve) in $1/M$ to the $B$ scale
(diamond or cross). The extrapolated points have been displaced
slightly for clarity.}
\label{fig:a3-one-over-M}
\end{figure}

The negative values of $A_3$ found from our simulation prompt us to
make two observations:
\begin{itemize}
\item
$1/M$ corrections to the heavy quark limit are large for $A_3$, as
shown in figure~\ref{fig:a3-one-over-M}. $A_3$ has the behaviour $A_3
\sim M A_2 \sim M^{3/2}$ in the limit $M\to\infty$, but in
extrapolating from scales around the charm mass, we start from a
region where the $A_1$ contribution, with behaviour $M A_1 \sim
M^{1/2}$, is non-negligible. $A_3$ exhibits the largest $1/M$
corrections we have found so far in several studies of processes
involving
heavy-to-heavy~\cite{ukqcd:fdfb,ukqcd:btodstar,ukqcd:lpllat94} or
heavy-to-light~\cite{ukqcd:hlff,ukqcd:btorho} quark transitions.  This
is confirmed by using $A_1$ and $A_2$ values extracted previously by
us~\cite{ukqcd:btorho} to calculate $A_3$ from the definition of
equation~(\ref{eq:a3def}): these calculated values are given in the
last column of table~\ref{tab:a0a3qsq} as $A_3^{\rm calc}$. We can
also reproduce the coefficients found for the $1/M$ expansion of $A_3
\Theta (m_B/M)^{3/2}$ by using the definition of
equation~(\ref{eq:a3def}) to express them in terms of the $1/M$
expansion coefficients of $A_1 \Theta (m_B/M)^{-1/2}$ and $A_2 \Theta
(m_B/M)^{1/2}$ found in reference~\cite{ukqcd:btorho}.
\item
Resonant contributions alone, meaning $q^2$ dependences of the form
given in equations~(\ref{eq:a0poleform2}) and~(\ref{eq:a3poleform}),
which satisfy the constraint $A_0(0) = A_3(0)$, are incompatible with
the measured difference in sign of $A_0$ and $A_3$.  The fact that
around $\qsqmax$, $A_0$ is positive and $A_3$ is negative, is a clear
failure of pole dominance models to explain the $q^2$ behaviour of the
form factors in a region away from the physical pole. Sum
rule~\cite{sumrules:cds} --\nocite{sumrules:abs,sumrules:ball}
\nocite{sumrules:bbd,sumrules:bkr,sumrules:cs} \cite{sumrules:cdp} and
quark model~\cite{stech,cybernetics} calculations, and the analysis of
reference~\cite{alop}, also indicate that simple single pole fits
cannot be made to all form factors.
\end{itemize}

The negative values found for $A_3$ near $\qsqmax$ are, in fact, still
compatible with heavy quark symmetry constraints as $M\to\infty$.  To
make a combined fit for $A_0$ and $A_3$, we use the following $q^2$
dependences\footnote{These are motivated by Stech's form factor
model~\protect\cite{stech} and by similar forms used recently by
L~Lellouch~\protect\cite{lpl:btopi-disp-rel} for the form factors
$f^+$ and $f^0$ in semileptonic $B\to\pi$ decay.}:
\begin{eqnarray}
A_0(q^2) &=& {\azero \over (1 - q^2/m_0^2)^{n_0}},
 \label{eq:a0model}\\
A_3(q^2) &=& \azero {(1 - c q^2/m_Bm_\rho)\over (1-q^2/m_3^2)^{n_0}}.
 \label{eq:a3model}
\end{eqnarray}
These forms satisfy the constraint $A_0(0) = A_3(0)$ and are
compatible with heavy quark symmetry provided that $\azero \sim
M^{-n_0+1/2}$ and $c$ is independent of $M$.

Because we had only four points for $A_3(q^2)$, covering a limited
range of $q^2$, we were unable to determine simultaneously the
parameters $\azero$, $c$, $m_0$ and $m_3$ in
equations~(\ref{eq:a0model}) and~(\ref{eq:a3model}). We therefore set
$m_3 = m_0$, as predicted by heavy quark symmetry in the infinite mass
limit. In fact we believe that the $1^+$ $B$ resonance should be
heavier than the $0^-$ resonance and therefore that $A_3$ should be
flatter in $q^2$ than $A_0$. By constraining both masses to be equal
we make an error apparently at quadratic order in the expansion of the
form factor $A_3$. However, the fit parameter $c$ can absorb quadratic
differences and our errors in fact begin at ${\cal O}(q^4)$.

This parameterisation fits our data well at the charm scale, as shown
in figure~\ref{fig:a0a3-c-and-b}a: the figure shows fitted curves for
$A_0$ and $A_3$ obtained for $n_0=1$ and $n_0 =5$. Over a limited
range of $q^2$, changes in $n_0$ can be compensated by variations in
the fitted mass parameter, $m_0$.

\begin{figure}
\vbox{%
\hbox to \hsize{\hfill
\epsfxsize=0.4\hsize\epsffile[45 39 509 503]{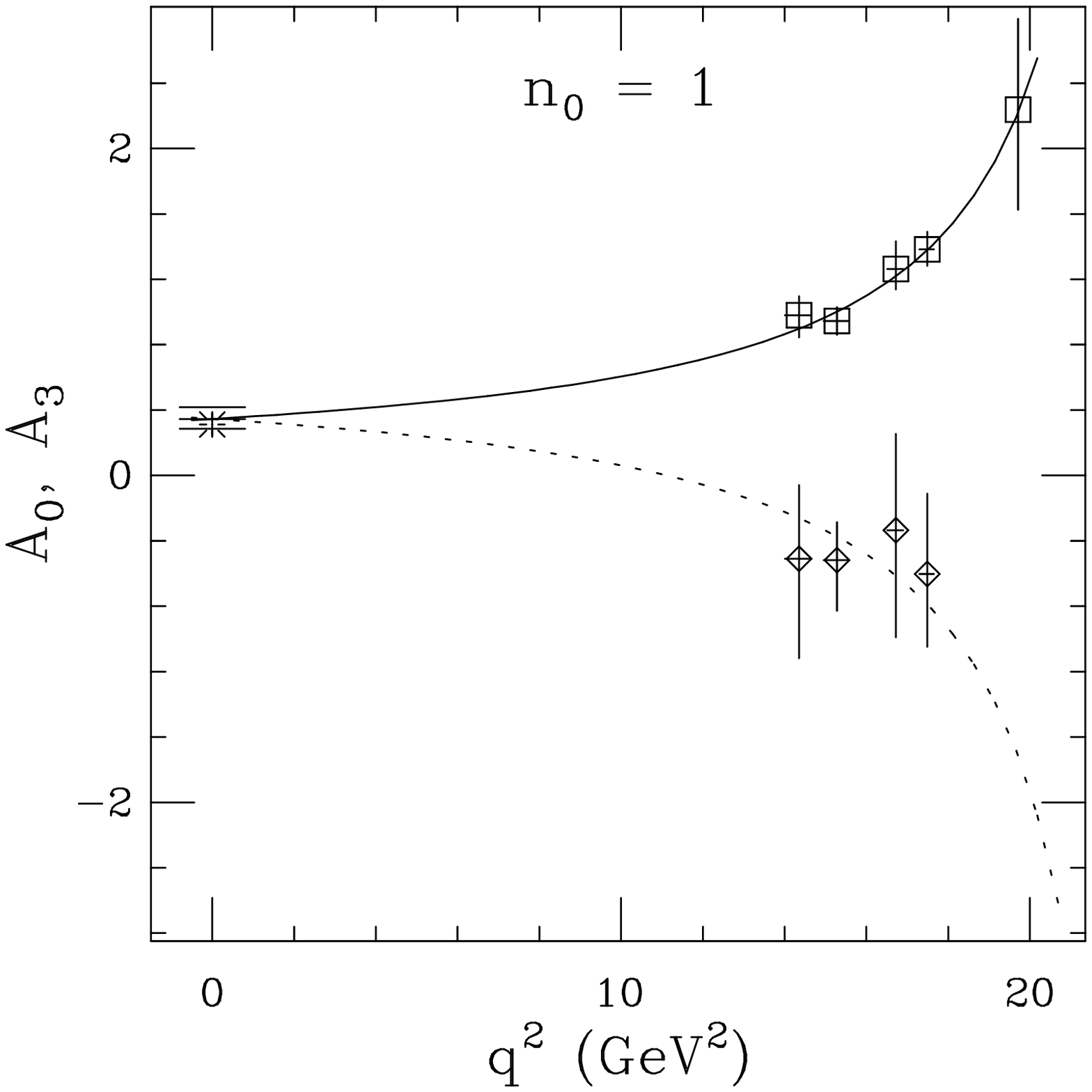}\hfill
\epsfxsize=0.4\hsize\epsffile[45 39 509 503]{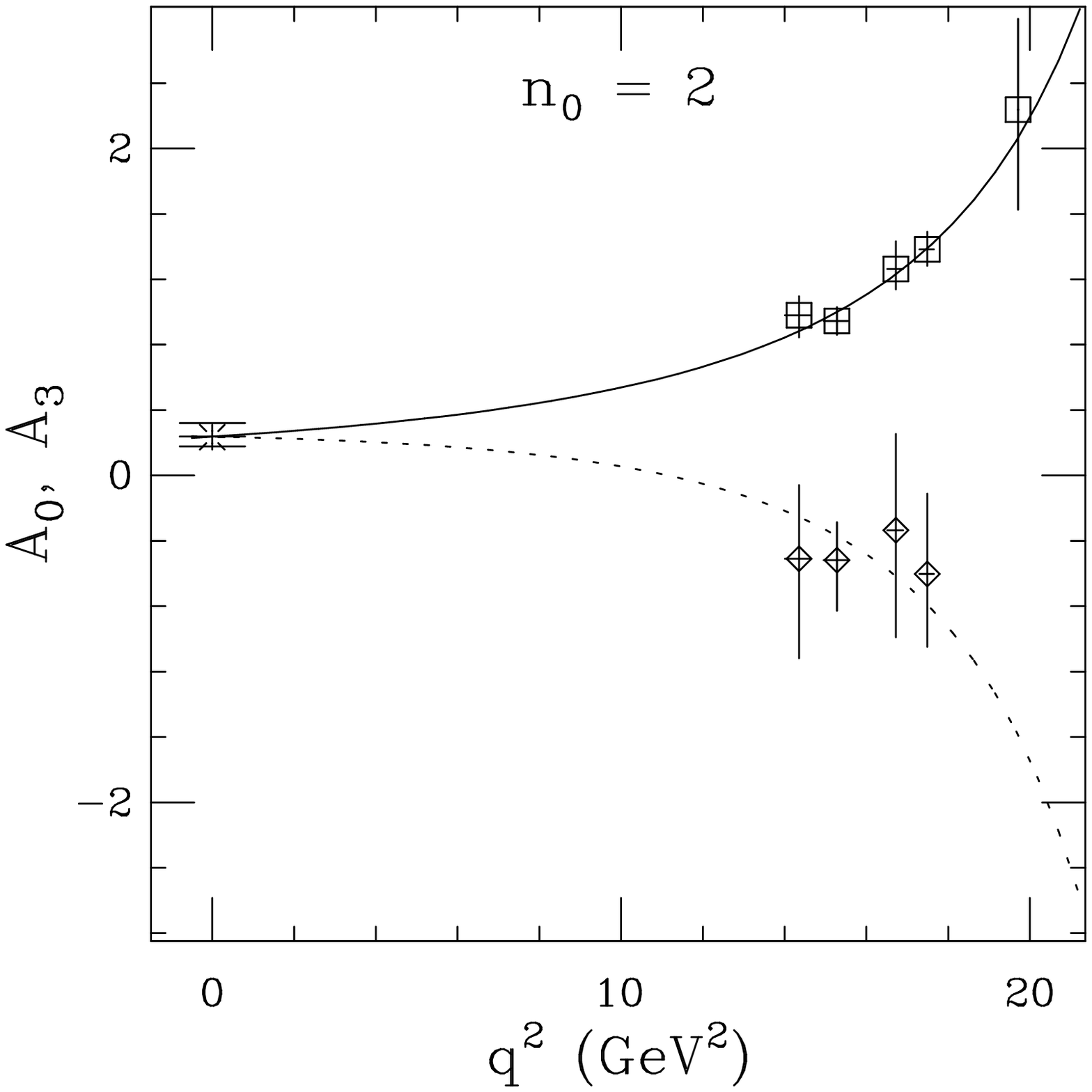}\hfill}
\medskip
\hbox to \hsize{\hfill
\epsfxsize=0.4\hsize\epsffile[45 39 509 503]{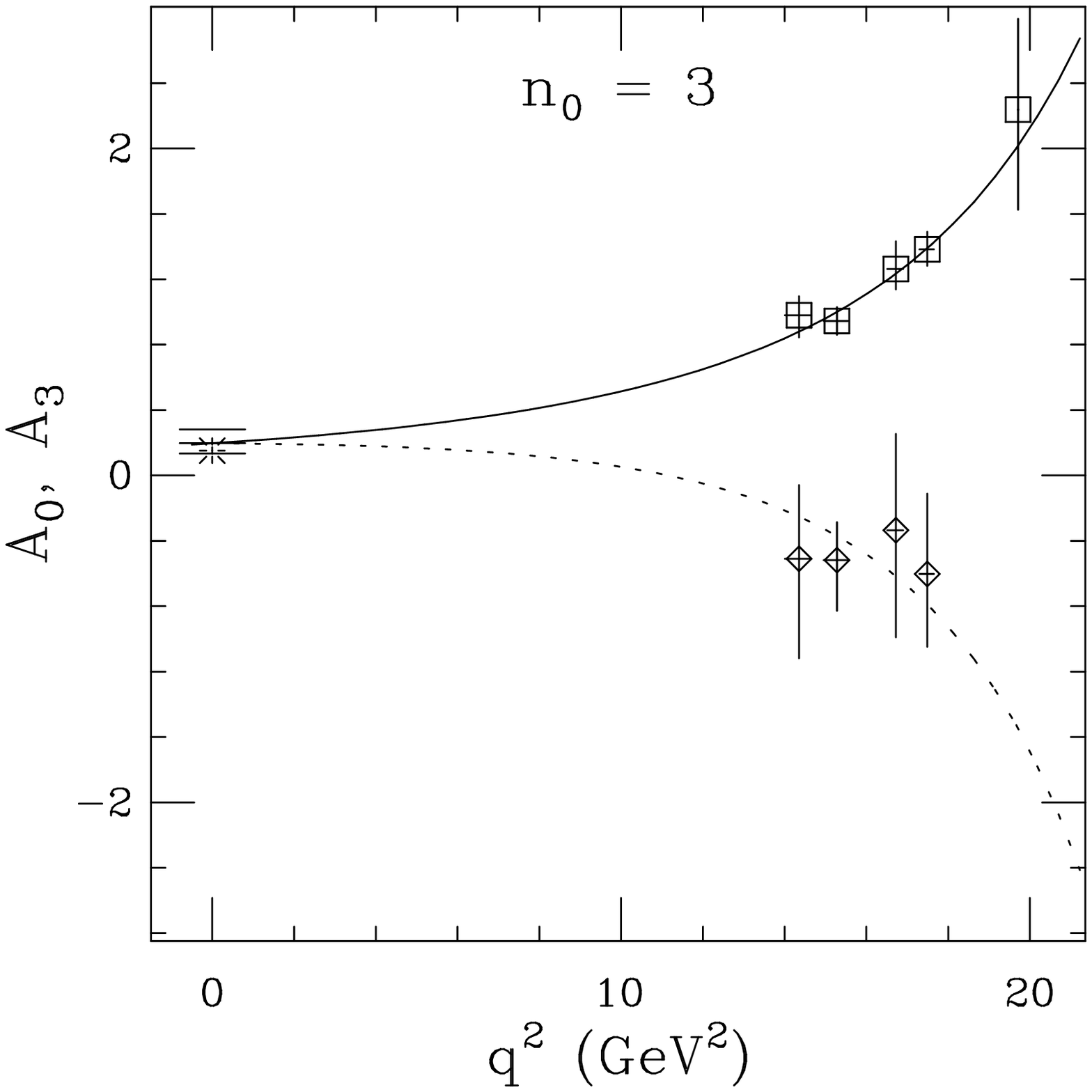}\hfill
\epsfxsize=0.4\hsize\epsffile[45 39 509 503]{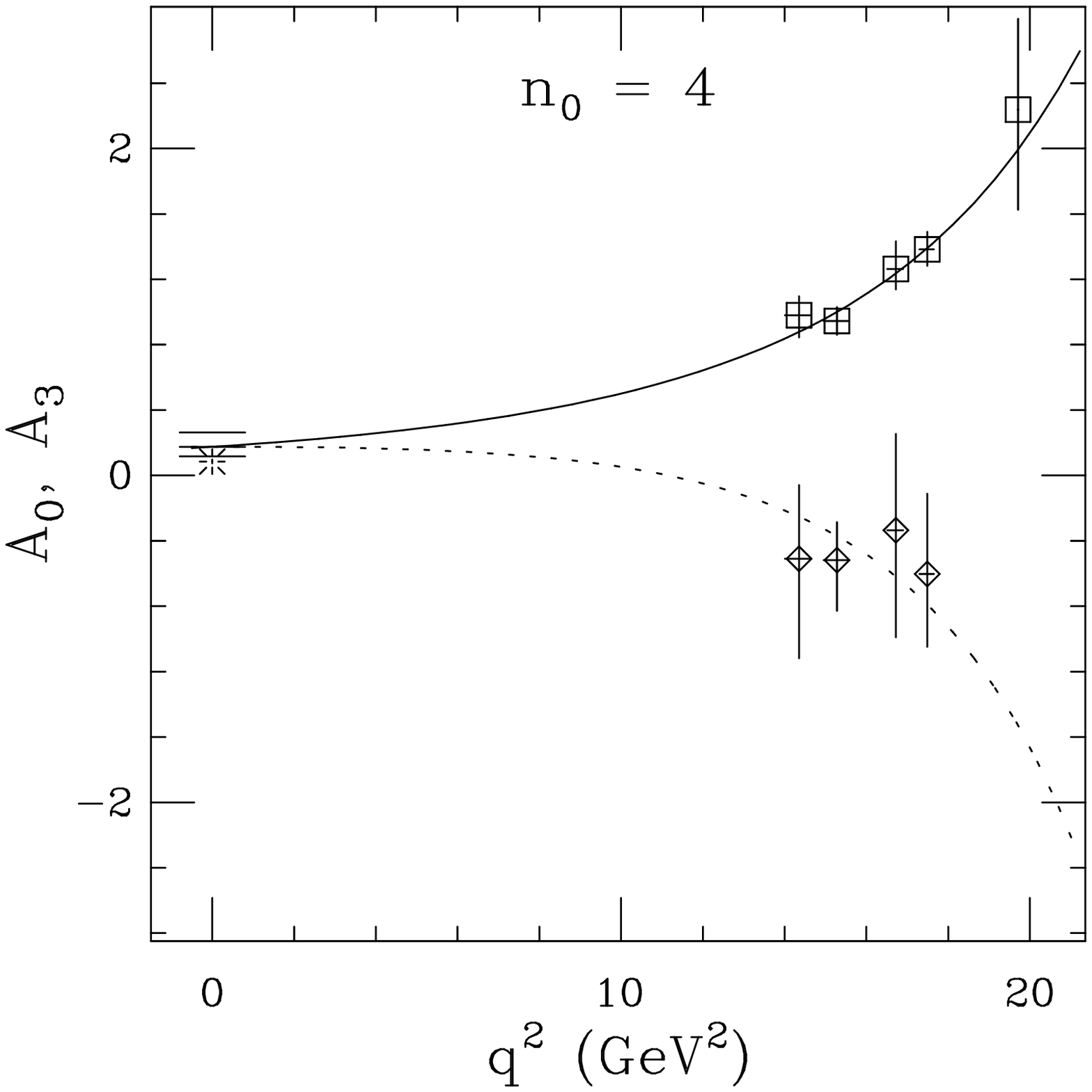}\hfill}
\medskip
\hbox to \hsize{\hfill
\epsfxsize=0.4\hsize\epsffile[45 39 509 503]{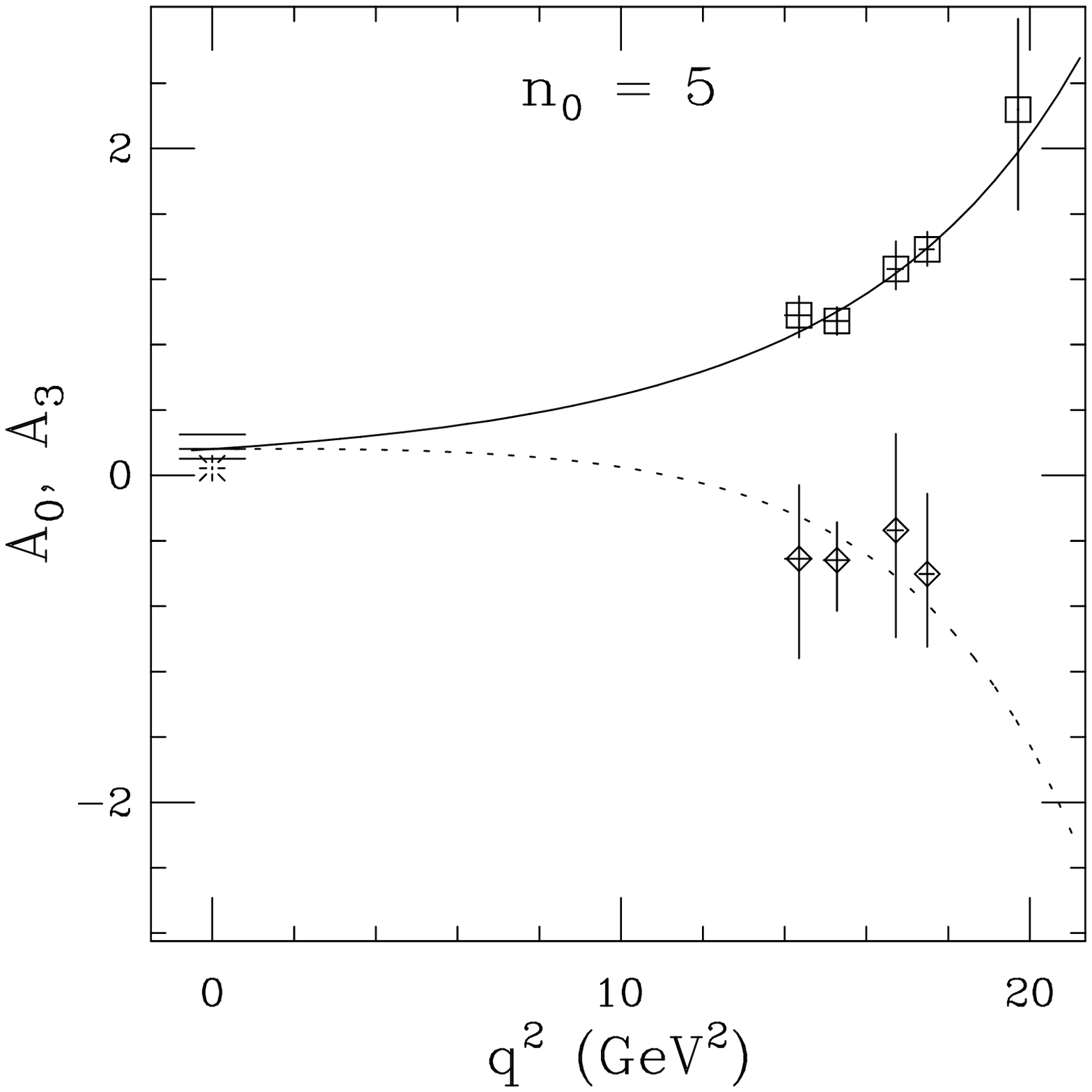}\hfill}
}
\caption[]{Combined fit to $A_0$ and $A_3$ at the $B$ scale using the
model of equations~(\protect\ref{eq:a0model})
and~(\protect\ref{eq:a3model}) for the solid and dotted curves
respectively, with different values of the polar power $n_0$ as
labelled on each plot. Squares are values for $A_0(q^2)$ and diamonds
are values for $A_3(q^2)$, the same in all five plots. The three
horizontal bars at $q^2=0$ show the value and errors for $\azero =
A_0(0) = A_3(0)$ in the combined fit (constant $\w$ method), and the
burst point shows $\azero$ determined from the $q^2=0$ extrapolation
method.}
\label{fig:a0a3fit-b}
\end{figure}

In figure~\ref{fig:a0a3fit-b} we show the results of the combined fit
at the $B$ scale for several choices of the polar power $n_0$ in the
model of equations~(\ref{eq:a0model}) and~(\ref{eq:a3model}). The
values obtained for $\azero$ are listed in table~\ref{tab:a0a3fit} as
one of the columns in the set labelled ``constant $\w$''. The table
also gives $\azero$ obtained from constrained fits for each of our
heavy quark mass values, followed by an extrapolation of $\azero$ to
the $B$ scale ($q^2=0$ extrapolation method), including terms up to
quadratic in the $1/M$ extrapolation. From the results of
table~\ref{tab:a0a3fit} we can estimate the error due to the
approximation $m_3 = m_0$.  The largest error corresponds to the
momentum channel $0\to1$, which has the largest value, $q^2 =
17.5(2)\gev$, used in the analysis for $A_3$. For this channel, taking
$m_3-m_0 \approx \Lambda_{\rm QCD} \approx 0.5 \gev$ (in the $D$ meson
system the analogous mass difference is $m_{D_1(1^+)} - m_{D^0(0^-)} =
2.423\gev - 1.865\gev = 0.558 \gev$), the error is 7\% for
$n_0=1,\ldots,5$.
\begin{table}
\hbox to\hsize{\def\arraystretch{1.2}\hss\begin{tabular}{c|c|cccc}
\hline
\tstrut
      & $q^2{=}0$ extrapolation &
 \multicolumn{4}{c}{constant $\w$} \\[-0.7ex]
$n_0$ & $\azero$ & $\azero$ & $c$ & $m_0/\!\gev$ & $\chi^2/{\rm dof}$
 \\[0.5ex]
\hline
\tstrut
1 & 0.31\er77 & 0.35\er76 & 0.37\er65 & 4.8\er21 & 0.31 \\
2 & 0.24\er44 & 0.24\er86 & 0.36\er65 & 5.5\er43 & 0.34 \\
3 & 0.15\er22 & 0.20\er86 & 0.36\er65 & 6.0\er63 & 0.36 \\
4 & 0.09\er11 & 0.18\er96 & 0.36\er65 & 6.6\er74 & 0.37 \\
5 & 0.04\er11 & 0.16\er96 & 0.36\er65 & 7.1\er85 & 0.38 \\
\hline
\end{tabular}\hss}
\caption[]{Values of the form factor $A_0(0)=A_3(0)$, together with
the fitted parameters $c$ and $m_0$, for $\btorho$ from a combined fit
of $A_0$ and $A_3$ using the model given in
equations~(\protect\ref{eq:a0model})
and~(\protect\ref{eq:a3model}). Errors are statistical only.}
\label{tab:a0a3fit}
\end{table}

In figure~\ref{fig:a0a3fit-b}, the $A_3$ curves drop steeply as $q^2$
approaches the pole at $m_3^2 = m_0^2$. This looks like an artefact of
using the same mass parameter for both $A_0$ and $A_3$, when we expect
$m_3$ to be heavier. Higher powers of $n_0$, while artificial,
effectively increase the pole mass and make the $q^2$ behaviour of
$A_3$ flatter away from the pole region.

We find a range $(0.16$--$0.35)\er96$ for $A_0(0)$. We do not include
a 10\% systematic error due to the chiral extrapolation and
discretisation effects, as discussed in reference~\cite{ukqcd:btorho},
because this error is much smaller than both the statistical and
systematic (owing to the different $q^2$-dependences assumed for $A_0$
and $A_3$) errors noted above.  Our range for $A_0(0)$,
$(0.16$--$0.35)\er96$, is compatible with the value obtained in
reference~\cite{sumrules:cds}: $A_0(0) = 0.24\pm0.02$.

The constrained fits give the same result as fitting $A_0$ alone
above, but having data for $A_3$ close to $\qsqmax$ can help choose
between different pole behaviours. Improved lattice simulations in the
near future with smaller statistical and systematic errors will
provide much better quality data for $A_3$ close to $\qsqmax$. This
will allow us to remove the restriction $m_3 = m_0$ and will help to
discriminate between the results for the different $q^2$ dependences
reported in table~\ref{tab:a0a3fit} and
figure~\ref{fig:a0a3fit-b}. For example, if $A_3(q^2)$ does not
decrease as $q^2\to\qsqmax$, then lower values of $A_0(0)$ will be
preferred.

In a previous paper~\cite{ukqcd:btorho}, we showed that $A_1(0)$ could
be extracted using an assumed pole form for its $q^2$ dependence. Here
we have extracted $A_0(0) = A_3(0)$, which, by the definition of $A_3$
in equation~(\ref{eq:a3def}), makes possible a determination of
$A_2(0)$. This can help constrain $q^2$ fits to $A_2$. Using a value
$A_1(0)=0.27\er74\er33$ taken from our earlier results in
reference~\cite{ukqcd:btorho}, we find\footnote{We quote $A_2(0)$
obtained from $A_3(0)$ with $n_0=2$. The central value of $A_2(0)$
ranges from $0.24$ for $n_0=1$ to $0.30$ for $n_0=5$. This range of
values is smaller than the size of the statistical errors quoted
above.} $A_2(0)=\Atwovalue$, where in both cases the first errors are
statistical and the second ones are systematic, due to discretisation
effects and uncertainties in the chiral extrapolation together with
the variation from changing $n_0$.  Since $A_2(\qsqmax) >
A_1(\qsqmax)$, we find that $A_2$ has a more pronounced $q^2$
dependence than $A_1$: this is consistent with sum
rule~\cite{sumrules:ball} and quark model~\cite{stech,cybernetics}
calculations, and with the analysis presented in~\cite{alop}. Our
result for $A_2(0)$ is the most precise obtained from the lattice to
date: it is compared with previous lattice determinations in
table~\ref{tab:a2values}.
\begin{table}
\hbox to\hsize{\hss
\begin{tabular}{ll}
\hline\tstrut
$A_2(0)$ & Reference \\[0.5ex]
\hline\tstrut
$\Atwovalue$ & this work \\
$0.24\pm0.56$ & APE ``a''~\protect\cite{ape:btopi} \\
$0.27\pm0.80$ & APE ``b''~\protect\cite{ape:btopi} \\
$0.38\pm0.18\pm0.04$ & ELC ``a''~\protect\cite{elc:btopi} \\
$0.49\pm0.21\pm0.05$ & ELC ``b''~\protect\cite{elc:btopi} \\
$0.72\er{35}{35}\er{10}7$ & \protect\cite{wup:slff}\\[0.5ex]
\hline
\end{tabular}\hss}
\caption[]{Lattice determinations of $A_2(0)$. The letters ``a'' and
``b'' denote results from two alternative methods of $1/M$
extrapolation (see~\protect\cite{ape:btopi}
and~\protect\cite{elc:btopi} for details).}
\label{tab:a2values}
\end{table}

Unfortunately, the quality of the numerical results for $A_0(0)$,
obtained here, and for $T_2$ in our previous work~\cite{ukqcd:hlff},
does not allow us to calculate the ratio $2T_2^{B\to K^*}(0)/A_0^{B\to
\rho}(0)$ of hadronic form factors in
equation~(\ref{eq:Roverdiffrate}).  In reference~\cite{ukqcd:hlff}, we
found two sets of values for $T_2(0)$, $0.15\er76$ and $0.26\er21$
depending on the $q^2$ behaviour (pole or constant, respectively)
assumed for the form factor $T_2(q^2)$. Some
authors~\cite{odonnell,sumrules:cds}, have suggested that the ratio
$I$ is close to one.  This, together with the results for $A_0$
presented here, would support low values for $T_2(0)$ and therefore a
pole-type behaviour for its $q^2$ dependence. In
reference~\cite{ukqcd:btorho}, by studying the ratio $A_1/2iT_2$
(predicted to be $1$ in the heavy-quark limit) and the form factor
$A_1$, we also found arguments supporting a pole-type behaviour for
$T_2$.

This study, combined with our previous studies of the form-factors
$A_1$, $A_2$ and $V$ in $\btorho$ decays~\cite{ukqcd:btorho}, $T1$ and
$T2$ in $\btokstargamma$ decays and $f^+$ and $f^0$ in $\btopi$
decays~\cite{ukqcd:hlff}, will enable us in the immediate future to
perform a global analysis of the heavy-to-light decays $\btopi$,
$\btokstargamma$ and $\btorho$.  Such a study will serve to check the
validity of different theoretical models, for example the one recently
suggested by Stech~\cite{stech}. In conjunction with theoretical
input, such as the constraints on $f^+$ and $f^0$ in $\btopi$ derived
from a dispersion relation analysis in~\cite{lpl:btopi-disp-rel}, it
could also help lattice calculations reduce the systematic
uncertainties related to the $q^2$ dependence of the form factors.

\section*{Conclusions}

We have determined a range of values for $A_0(0)$ from the lattice for
the first time. The range we find is: $(0.16$--$0.35)\er96$. As data
on both $A_0$ and $A_3$ improves, it could become an important
ingredient in the determination of $\vub$.

From fits to $A_0$ alone, it is hard to dismiss high values for
$A_0(0)$ arising from an assumed single pole behaviour of the form
factor. Information from $A_3$, although not of high quality, can help
determine $A_0(0)$ by disfavouring such high values. With our present
data we cannot rule out the higher values, but with better quality
data, in particular $A_3(q^2)$ values with $q^2$ closer to $\qsqmax$,
a constrained fit to $A_0$ and $A_3$ could give more reliable answers
(as was found for the form factors $f^+$ and $f^0$ for $\btopi$
in~\cite{ukqcd:hlff}). This would allow us to reduce the range of
values quoted for $A_0(0)$.

$A_3$ itself, for $q^2$ near $\qsqmax$, has a large dependence on the
heavy quark mass, changing from positive to negative as one scales
from the charm mass to the bottom mass. Because $A_3(q^2)$ at the $B$
scale is negative when $A_0(q^2)$ is positive at the same $q^2$, close
to $\qsqmax$, it cannot be fitted solely by resonant terms. Hence, our
lattice measurements bear out the predictions of quark models.

Combining the results for $A_0(0)$ with earlier results for $A_1$
allows us to determine $A_2(0) = \Atwovalue$. Combining this
with improved lattice measurements of $A_2(q^2)$ for $q^2$ near
$\qsqmax$ could allow the determination of the full $q^2$ dependence
of $A_2$.

Future lattice simulations, with improved statistical and systematic
errors, would allow us to take $m_3 \neq m_0$ in the fits to
equations~(\ref{eq:a0model}) and~(\ref{eq:a3model}).  We believe that
in these circumstances the most physically motivated polar power,
$n_0=1$, would be preferred. We expect that the results obtained would
coincide with those presented here, where we have set $m_3=m_0$ (as
predicted by heavy quark symmetry) and have limited ourselves to a
$q^2$-region away from the pole where higher polar powers effectively
increase the pole mass and make the $q^2$ behaviour of $A_3$ flatter.

\subsection*{Acknowledgements}

We thank Laurent Lellouch, Chris Sachrajda and Hubert Simma for
comments and other members of the UKQCD collaboration for the original
calculations of the lattice correlation functions.
JN thanks the Theory Group in the Southampton Physics Department for
its kind hospitality during the early stages of this project.
This research was supported by the UK Science and Engineering Research
Council under grants GR/G 32779 and GR/H 49191, by the
University of Edinburgh and by Meiko Limited.
We are grateful to Edinburgh University Computing Service and, in
particular, to Mike Brown, for maintaining service on the Meiko i860
Computing Surface.
%
%
We acknowledge the Particle Physics and Astronomy Research Council for
travel support under grant GR/J 98202.


\begin{thebibliography}{99}
\bibitem{bsw}M~Wirbel, B~Stech and M~Bauer, Z. Phys. {\bf C29}
  (1985) 637; M~Bauer and M~Wirbel, Z. Phys. {\bf C42} (1989) 671
\bibitem{isgurwise:hqet}N~Isgur and M~B~Wise, Phys. Rev. {\bf D42}
  (1990) 2388
\bibitem{neubert:physrep}M~Neubert, Phys. Rept. {\bf 245} (1994) 259
\bibitem{odonnell}P~J~O'Donnell and Q~P~Xu, Phys. Lett. {\bf
  B325} (1994) 219; P~J~O'Donnell and H~K~K~Tung, Phys. Rev. {\bf D48}
  (1993) 2145
\bibitem{santorelli}P~Santorelli, Z. Phys. {\bf C61} (1994) 449
\bibitem{cleo:btokstargamma}CLEO collaboration, R~Ammar {\em et al.},
  Phys. Rev. Lett. {\bf 71} (1993) 674
\bibitem{LD:atwoodetal}D.~Atwood, B.~Blok and A.~Soni, SLAC preprint
  SLAC--PUB--95--6635, Brookhaven preprint BNL--60709, Technion preprint
  PH--94--11, hep-ph/9408373 (1994)
\bibitem{longdist}E.~Golowich and S.~Pakvasa, Phys. Lett. B. {\bf
  205}, 393 (1988); Phys. Rev. D {\bf 51} (1995) 1215
\bibitem{sumrules:cds}P~Colangelo, F~De~Fazio and P~Santorelli,
  Phys. Rev. {\bf D51} (1995) 2237
\bibitem{sumrules:abs}A~Ali, V~M~Braun and H~Simma, Z. Phys. {\bf
  C63} (1994) 437
\bibitem{sumrules:ball}P~Ball, Phys. Rev. {\bf D48} (1993) 3190
\bibitem{sumrules:bbd}P~Ball, V~M~Braun and H~G~Dosch, Phys.~Rev.
  {\bf D44} (1991) 3567
\bibitem{sumrules:bkr}V~M~Belyaev, A~Khodjamirian and R~R\"uckl,
  Z. Phys. {\bf C60} (1993) 349 and Phys. Rev. {\bf D51} (1995) 6177
\bibitem{sumrules:cs}P~Colangelo and P~Santorelli, Phys. Lett. B {\bf
  327} (1994) 123
\bibitem{sumrules:cdp}P~Colangelo, C~A~Dominguez and N~Paver,
  Phys. Lett. B {\bf 352} (1995) 134
\bibitem{stech}B~Stech, Phys. Lett. B {\bf 354} (1995) 447
\bibitem{cybernetics}R~N~Faustov, V~O~Galkin and A~Yu~Mishurov,
  Phys. Lett. B {\bf 356} (1995) 516 and Cybernetics Council,
  Moscow, preprint hep-ph/9508262
\bibitem{alop}R~Aleksan {\em et al.}, Phys. Rev. {\bf D51} (1995)
  6235
\bibitem{cleo:exclusive}M~Selen, talk given at 6th International
  Symposium on Heavy Flavor Physics, Pisa, Italy, June 1995,
  hep-ph/9508304; L~K~Gibbons, private communication
\bibitem{ape:btopi}APE collaboration, C~R~Allton {\em et al.},
  Phys. Lett. B {\bf 345} (1995) 513
\bibitem{elc:btopi}As~Abada {\em et al.}, Nucl. Phys. {\bf B416}
  (1994) 675
\bibitem{wup:slff}S~G\"usken, G~Siegert and K~Schilling, talk
  presented at Japan-Germany Seminar on QCD on Massively           
  Parallel Computers, Yamagata, Japan, March 1995,
  Wuppertal/J\"ulich preprint, HLRZ 95--38, WUB 95--22,
  hep-lat/9507002
\bibitem{sw-action}B~Sheikholeslami and R~Wohlert, Nucl. Phys. {\bf
  B259} (1985) 572
\bibitem{heatlie:clover-action}G~Heatlie {\em et~al.}, Nucl. Phys.
  {\bf B352} (1991) 266
\bibitem{ukqcd:strange-prd}UKQCD collaboration, C~Allton {\em
  et~al.}, Phys. Rev. {\bf D49} (1994) 474
\bibitem{ukqcd:static}UKQCD collaboration, A~K~Ewing {\em et~al.},
  Edinburgh preprint 95/550, Southampton preprint SHEP--95--20,
  Swansea preprint SWAT/78
\bibitem{zva}A~Borrelli {\em et al.}, Nucl.~Phys.\ {\bf  B373}
  (1992) 781
\bibitem{ukqcd:dtok}UKQCD collaboration, K~C~Bowler {\em et~al.},
  Phys. Rev. {\bf D51} (1995) 4905
\bibitem{ukqcd:hlff}UKQCD collaboration, D~R~Burford {\em et al.},
  Nucl. Phys. {\bf B447} (1995) 425
\bibitem{ukqcd:btorho}UKQCD collaboration, J~M~Flynn {\em et al.},
  Southampton preprint SHEP--95--18, hep-ph/9506398, Nucl. Phys. B
  in press
\bibitem{ukqcd:fdfb}UKQCD collaboration, R~M~Baxter {\em et al.}, Phys.
  Rev. {\bf D49} (1994) 1594
\bibitem{ukqcd:btodstar}UKQCD collaboration, S~P~Booth {\em et al.},
  Phys. Rev. Lett. {\bf 72} (1994) 462; UKQCD collaboration, K~C~Bowler
  {\em et al.}, hep-ph/9504231, Phys. Rev. {\bf D} in press
\bibitem{ukqcd:lpllat94}L~P~Lellouch in Proc. LATTICE 94: 12th
  International Symposium on Lattice Field Theory, Bielefeld,
  Germany, Eds. F~Karsch {\em et al.}, Nucl. Phys. B (Proc. Suppl.)
  {\bf 42} (1995) 421; Marseille preprint CPT--95/P.3196, to appear in
  Proc. XXX Rencontres de Moriond, Les Arcs, France, March 1995
\bibitem{lpl:btopi-disp-rel}L~Lellouch, Centre de Physique
  Th\'eorique, Marseille preprint, CPT--95/P.3236, hep-ph/9509358
  (1995)
\end{thebibliography}
\end{document}